\documentclass[preprint2]{aastex6}

\usepackage{mathrsfs}
\usepackage{amsmath}
\usepackage{booktabs}
\usepackage{longtable}

\DeclareMathOperator{\sinc}{sinc}

\shorttitle{First results from SONG}
\shortauthors{Grundahl et al.}

\begin{document}
\title{First results from the Hertzsprung SONG Telescope:\\Asteroseismology of the G5 subgiant star $\mu\,$Her
\footnote{Based on observations made with the Hertzsprung SONG telescope operated on the Spanish
Observatorio del Teide on the island of Tenerife by the Aarhus and Copenhagen Universities and by the 
Instituto de Astrof\'isica de Canarias.} }
\altaffiltext{1}{Stellar Astrophysics Centre (SAC), Department of Physics and Astronomy, Aarhus University Ny Munkegade 120, DK-8000}
\altaffiltext{2}{Sydney Institute for Astronomy, School of Physics, University of Sydney 2006, Australia}
\altaffiltext{3}{Instituto de Astrof\'isica de Canarias, E-38205 La Laguna, Tenerife, Spain}
\altaffiltext{4}{Universidad de La Laguna, Dpto. Astrofísica, E-38206 La Laguna, Tenerife, Spain}
\altaffiltext{5}{Niels Bohr Institute, University of Copenhagen, Juliane Maries vej 30, DK-2100 Copenhagen, Denmark}
\altaffiltext{6}{School of Physics and Astronomy, University of Birmingham, Edgbaston, Birmingham B15 2TT, UK}
\altaffiltext{7}{Niels Bohr Institute and Centre for Star and Planet Formation, University of Copenhagen, {\O}ster Voldgade 5, 1350 Copenhagen K, Denmark}

\altaffiltext{8}{Zentrum f{\"u}r Astronomie der Universit{\"a}t Heidelberg (ZAH), Institut f{\"u}r Theoretische Astrophysik, Albert-Ueberle-Str. 2, 69120 Heidelberg, Germany}
\altaffiltext{9}{Department of Geoscience, Aarhus University, H{\o}egh-Guldbergs Gade 2, DK-8000 Aarhus C, Denmark}
\altaffiltext{10}{Zentrum f\"ur Astronomie der Universit\"at Heidelberg, Landessternwarte, K\"onigstuhl 12, 69117 Heidelberg}
\altaffiltext{11}{Centre for Electronic Imaging, The Open University, Milton Keynes MK7 6AA, UK}

\author{F. Grundahl                \altaffilmark{1}}
\author{M. Fredslund Andersen      \altaffilmark{1} }
\author{J. Christensen-Dalsgaard   \altaffilmark{1} }
\author{V. Antoci                  \altaffilmark{1} }
\author{H. Kjeldsen                \altaffilmark{1} }
\author{R. Handberg                \altaffilmark{1} }
\author{G. Houdek                  \altaffilmark{1} }
\author{T. R. Bedding              \altaffilmark{2, 1} }
\author{P. L. Pall\'e              \altaffilmark{3, 4} }
\author{J. Jessen-Hansen           \altaffilmark{1} }
\author{V. Silva Aguirre           \altaffilmark{1} }
\author{T. R. White                \altaffilmark{1} }
\author{S. Frandsen                \altaffilmark{1} }
\author{S. Albrecht                \altaffilmark{1} }
\author{M. I. Andersen             \altaffilmark{5} }
\author{T. Arentoft                \altaffilmark{1} }
\author{K. Brogaard                \altaffilmark{1} }
\author{W. J. Chaplin              \altaffilmark{6,1} }
\author{K. Harps\o e               \altaffilmark{7} }
\author{U. G. J\o rgensen          \altaffilmark{7} }
\author{I. Karovicova              \altaffilmark{8} }
\author{C. Karoff                  \altaffilmark{1, 9} }
\author{P. Kj\ae rgaard Rasmussen  \altaffilmark{5} }
\author{M. N. Lund                 \altaffilmark{6, 1} }
\author{M. Sloth Lundkvist         \altaffilmark{10, 1} }
\author{J. Skottfelt               \altaffilmark{7, 11} }
\author{A. Norup S\o rensen        \altaffilmark{5} }
\author{R. Tronsgaard              \altaffilmark{1} }
\and
\author{E. Weiss                   \altaffilmark{1} }

\email{fgj@phys.au.dk}


\begin{abstract}

We report the first asteroseismic results obtained with the Hertzsprung SONG Telescope 
from an extensive high-precision radial-velocity observing campaign of the subgiant 
$\mu\,$Herculis.  The data set was collected {during} 215 nights in 2014 and 2015.  We detected 
a total of 49 oscillation modes with $l$ values from 0 to 3, including {some} $l=1$ mixed modes. 
Based on the rotational splitting observed in $l=1$ modes, we determine a rotational period of 
52 days and a stellar inclination angle of 63 degrees. The parameters obtained through modeling 
of the observed oscillation frequencies agree very well with independent observations and imply 
a stellar mass between 1.11 and  1.15\,M$_\odot$ and an age of $7.8^{+0.3}_{-0.4}$\,Gyr. 
Furthermore, the high-quality data allowed us to determine the acoustic depths of the \ion{He}{2} 
ionization layer and the base of the convection zone.

\end{abstract}





\section{Introduction}

Asteroseismology of solar-like oscillations has blossomed as an
observational science in the past few years, thanks to the steady flow of
high-precision photometry from the CoRoT and {\em Kepler} space missions
\citep[see][for a review]{Chaplin2013}.  From {\em Kepler} we now have
oscillation spectra, based on four years of continuous observations, for
hundreds of main-sequence stars and tens of thousands of red giants.
However, ground-based spectroscopic measurements of solar-like oscillations
\citep[see][and references therein for a review]{Bedding2012} still have 
an important role to play.  
They can be used to target specific stars of interest anywhere in
the sky, and they can provide a higher signal-to-noise ratio than photometry
because the stellar background from granulation is much lower in velocity
than intensity compared to the oscillations \citep[see, e.g.,][Fig.\,1]{Grundahl2007}.  
This property makes the detection of lower frequency and $l=3$ modes less difficult.
Subgiants are particularly interesting for
asteroseismology because some of their oscillations occur as mixed modes,
which have characteristics of both pressure and gravity modes and are very
sensitive to the conditions in the stellar core \citep{Christ1995}.

SONG (Stellar Observations Network Group) is planned as a network of 1-m
telescopes that will carry out high-precision radial-velocity measurements of
stars.  The first node at Observatorio del Teide on Tenerife has been
operating since 2014 and consists of the Hertzsprung SONG Telescope, which
is equipped with a coud\'e \'echelle spectrograph with an iodine cell \citep{Grundahl2007}. 
Here, we present observations over two observing
seasons (2014 and 2015) of the G5 subgiant star $\mu\,$Herculis ($\mu$\,Her).
Importantly, this star turns out to have a frequency spacing that is highly
favorable for single-site observations \citep{Arentoft2014}.  Our observations span a total of
215 nights and have yielded an oscillation spectrum with high
signal-to-noise ratio and high frequency resolution, allowing the most detailed
asteroseismic study ever performed for a subgiant observed from ground.


\section{Basic properties of $\mu$\,Her}

The star $\mu$\,Her (HD 161797, HR 6623, HIP 86974) is a bright G5 subgiant.
Solar-like oscillations were detected by \cite{Bonanno2008}
using iodine-referenced radial--velocity observations over seven nights with the
3.6-m Italian TNG Telescope on La Palma. They detected a clear excess of
power centered at a frequency of $\nu_{\rm max} = 1200\,\mu$Hz and found the
most likely value for the large frequency separation to be $\Delta\nu =
56.5\,\mu$Hz.  Based on this value, \cite{Bonanno2008} extracted frequencies 
for 15 individual oscillation modes, which were {subsequently} used for 
theoretical modeling \citep{Pinheiro2010, Yang2010}.  

In the following sections we discuss the fundamental stellar properties of
$\mu\,$Her as input for modeling the measured oscillation frequencies.  
Estimates for the radius and luminosity are also provided for later comparison 
to the model results.

\subsection{T$_{\rm eff}$, $\log g$ and {\rm [Fe/H]} } 

Because of its brightness, the basic parameters for $\mu\,$Her have been 
determined in many studies.  The most recent publication is the 2016 
version of the PASTEL catalog \citep{Soubiran2016}, which also summarizes 
nearly all literature values.

The reported effective temperature determinations range from 5397
\footnote{ {\cite{Baines2014} provided a $T_{\rm eff}$ 
estimate of 5317\,K based on an angular--diameter measurement. 
The low value $T_{\rm eff}$ is probably due to a low value for
the estimated bolometric flux, with a reported uncertainty which 
appears to be unrealistically small.  Note also that the reported 
parallaxes for $\mu$\,Her and HD\,188512 in their Table 2 are incorrect.} }
to 5650\,K, $\log g$ from 3.7 to 4.1 and [Fe/H] values between +0.04 and +0.3. 
Most of these studies employed `standard' 1D--LTE analysis of high-resolution,  
high signal-to-noise spectra to determine these parameters and arrived slightly
different conclusions.  We do 
not have a quantitative way to decide which values are the best.  We 
therefore adopted the most recent parameters \cite[hereafter J15]{Jofre2015} 
and list them in Table~\ref{tab:1}. To reflect that this choice is a compromise, we 
assigned larger uncertainties than reported by J15.  Specifically, \cite{Bruntt2010}
have discussed the accuracy of the determination of stellar temperature, 
gravity and [Fe/H] and concluded that realistic error bars for these 
quantities are 80\,K, 0.08\,dex, and 0.07\,dex, respectively. We have adopted 
these values here.  Finally, J15 also determined $v\,\sin i = 1.7$\,km\,s$^{-1}$, 
which is in accordance with expectations for an old, slightly evolved low--mass star. 

\subsection{Luminosity and radius} 

To estimate the luminosity we used the measured $V\,=\,3.42$ \citep{Bessel2000}, the 
{\em Hipparcos} parallax (120.33$\pm$0.16\,mas), and assumed $A_V\,=\,0$, 
which yielded $M_V=3.82$. 
The bolometric correction was calculated using equation 9 from \cite{Torres2010}. 
We used the \cite{Casagrande2014} 
 $V$ filter bolometric corrections and found $-$0\fm086 and $-$0\fm068 
for $\mu$\,Her and the Sun, respectively. 
Based on these values, we determined 
$L\,=\,2.54\pm0.08\,{\rm L_{\odot}}$. 
The radius can be determined from angular--diameter measurements. Observations of $\mu\,$Her 
were recently made with the Precision Astronomical Visual Observations (PAVO) beam 
combiner \citep{Ireland2008} at the Center for High Angular Resolution Astronomy (CHARA) 
Array \citep{tenBrummelaar2005}. A fit of a uniform-disc model to these observations 
resulted in a uniform-disc diameter of $\theta_{\rm UD}\,=\,1.821\pm0.018$\,mas 
(Karovicova et al., in prep.). We determined a linear limb-darkening coefficient in 
the $R$ band (0.60$\pm$0.04) by interpolating the model grids of \citet{Claret2011} 
to the spectroscopic values of $T_\mathrm{eff}$, $\log g$ and [Fe/H]. The subsequent 
limb-darkened diameter is determined to be $\theta_{\rm LD}\,=\,1.93\pm0.02$\,mas. 
Using the parallax, this translates to a radius of $R\,=\,1.73\pm0.02 {\rm R_{\odot}}$. 

While angular diameters are often used to determine effective stellar temperatures
we have opted not to do this here because we have found three independent literature
values for the bolometric flux that differ by 25\%, which makes it problematic to 
select the correct value \citep{Mozurkewich2003, Boyajian2013, Baines2014}. 
We note that if we adopt the luminosity from photometry,  
our interferometric radius and neglecting the uncertainty in the parallax, the 
inferred temperature for $\mu\,$Her becomes 5540$\pm80$\,K, which is fully consistent 
with the adopted spectroscopic temperature.

\subsection{Activity} 

There are only two reports on the activity level for $\mu$\,Her, based 
on the Ca\,HK lines: \cite{Wright2004} reported $\log\,$R$_{\rm HK} = -5.11$ 
and \cite{Isaacson2010} found $\log\,$R$_{\rm HK} = -5.08$.  Both values 
are lower than the level found for the Sun, suggesting that $\mu$\,Her is 
a rather inactive star, consistent with its evolutionary stage.  This is, 
however, contradicted by the newly released measurements by the Mount 
Wilson Observatory HK Project\footnote{\url{http://www.nso.edu/node/1335}}. 
These measurements of $\mu$\,Her indicate an abrupt change in the S index from 
0.14 to more than 0.3. At this stage, it is impossible to conclude 
whether this jump is of stellar origin, and only additional data can solve
this ambiguity.

\subsection{Multiplicity}\label{subsec:binary} 

\cite{Roberts2016} provided a detailed summary of the quadruple nature of $\mu\,$Her.
All other components of the system are M-type dwarfs. 
Interestingly, the inner pair ($\mu\,$Her and one of the M-dwarfs) of the system 
has an orbital inclination 
of $63\,\pm\,5$ degrees \citep{Roberts2016}, which agrees very well with the 
inclination of the $\mu\,$Her rotation axis determined from our seismic measurements (see 
Sec.~\ref{sec:discussion}).  
From the analysis of published radial--velocity and astrometric measurements, \cite{Roberts2016} 
determined an orbital period of $\sim$\,100 years and 
concluded that this pair is currently close to the lower 
inflection point of the radial--velocity curve.  
We expect to cover this portion of the 
orbit with SONG radial--velocity measurements in the coming years. 

\begin{table*}[h]
\begin{center}
\caption{Classical parameters for $\mu\,$Her.\smallskip\label{tab:muher-parameters}}
\begin{tabular}{lccl}
\tableline\tableline
Parameter                      & Value  & Uncertainty & Reference\\
\tableline 
$T_{\rm eff}$\,[K]             & 5560   & 80     & J15, our {uncertainty} \\
${\rm [Fe/H]}$\,[dex]          & 0.28   & 0.07   & J15, our {uncertainty} \\
$\log g$\,[dex]                & 3.98   & 0.10   & J15, our {uncertainty} \\
$v\,sin\,i$\,[km\,s$^{-1}$]    & 1.7    & 0.4    & J15               \\
Parallax [mas]                 & 120.33 & 0.16   & \cite{vanLeeuwen2007} \\
$\theta_{\rm LD}$[mas]         & 1.93   & 0.03   & Derived here \\
$R / {\rm R_{\odot}}$          & 1.73   & 0.02   & Derived here \\
$L / {\rm L_{\odot}}$          & 2.54   & 0.08   & Derived here \\
$M_V$                          & 3.82   & 0.03   & Derived here \\
System velocity [km\,s$^{-1}$] & -17.07 & 0.12   & SIMBAD\\
log$R'_{\rm HK}$               & $-5.1$ & 0.1    & \cite{Isaacson2010}\\
       
\tableline
\end{tabular}
\end{center}
\label{tab:1}
\end{table*}


\section{The SONG prototype and observations of $\mu\,$Her}

$\mu\,$Her was observed with the automated 1-m Hertzsprung SONG telescope 
\citep[][]{Andersen2014} 
at Observatorio del Teide during the summers of 2014 (105 nights) and 2015 
(110 nights).  All spectra for radial--velocity determination were 
collected through an iodine cell for precise wavelength calibration. 
Each observation consisted of a 120\,s exposure, with 2.3\,s readout time 
for the CCD camera. 
A spectral resolution of 90,000 was used throughout the entire observing campaign.  
The median count per pixel at 5560\,{\AA} was 25514\,ADU.
The spectra have 51 spectral orders covering 4400\,{\AA} to 6900\,{\AA}.
A total of nearly 30000 spectra were collected
during the two observing seasons. All the 2014 spectra were reduced using 
an IDL-based pipeline that uses the routines of \cite{Piskunov2002}. 
For the 2015, data the extraction pipeline was based on the C++ 
re-implementation of the same routines by \cite{Ritter2014}. 
Before each observing night, calibration frames (bias {frames}, flat fields and 
ThAr spectra) were obtained and applied nightly. The extracted spectra,
with the superimposed iodine absorption spectrum, were
analysed with the code {\tt iSONG} \citep[e.g.,][]{Corsaro2012, Antoci2013}.  
This code closely follows the procedures outlined 
by \cite{Butler1996} to extract the stellar radial velocities. 
To generate the required intrinsic stellar template, the bright fast-rotating
star HR\,6410 was observed at $R=110,000$ to determine the spectral-line-spread 
function of the spectrograph. This was used to deconvolve a high-S/N
spectrum of $\mu\,$Her obtained without the iodine cell.
{For each spectrum the} RV code extracted velocities in 24 spectral 
orders, each subdivided into 22
``chunks'' of 91 pixels {(approximately 2{\AA}) }. This resulted in 528 
independent radial--velocity estimates.
We calculated the final velocities {as the weighted average velocity of all chunks.}
{The noise was estimated} from  the power--spectrum analysis in
Section~\ref{sec:processing-time-series}, resulting in an average precision 
of $\sim$1.5\,m\,s$^{-1}$ per spectrum.  For each exposure, we calculated 
the barycentric Julian mid-time and barycentric velocity correction using 
the program {\tt BarCor}\footnote{\url{sirrah.troja.mff.cuni.cz/\~ mary} }
by M. Hrudkov{\'a}. 


\section{Initial processing of the time series} \label{sec:processing-time-series}

\begin{figure}[tbh]
\epsscale{1.00}
\plotone {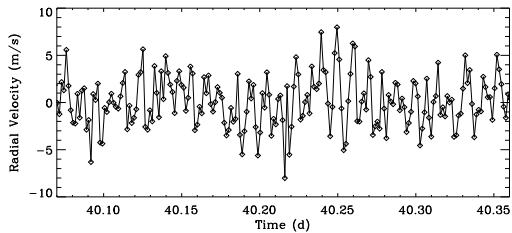}
\caption{Seven hours of raw $\mu\,$Her velocity data for one of the best nights 
in 2014.  The high quality of the data allows us to directly see the oscillations 
in the time-series data.}
\label{fig:time-series}
\end{figure}

The quality of the data is very high (a 7-hr segment {from one of the 
best nights} of the time series is shown in Fig.~\ref{fig:time-series}).  
However, the data 
quality does vary slightly from night to night, and also within 
nights, as a function of zenith distance, seeing, and instrumental
effects. In order to optimize the signal-to-noise ratio in the power spectrum, we
estimated the statistical quality of each measurement.  To do this, we 
first created a high-pass-filtered version by smoothing the time series 
with a Gaussian filter with a FWHM of 500\,s, which {was} then subtracted from 
the original data to remove all p-mode oscillations and long-term drifts. 
This filtered time series was used to estimate the local 
variance,~$\sigma_i^2$, which we calculated as the moving mean of the square 
over a duration of 6 hours (about 180 data points). In this way 
only slow changes were included in the estimates of the local variance.
Data points that deviated more than 4 times the local root-mean-square (rms) scatter were
removed from the raw time series and from the high-pass filtered series. 
{This} 4-$\sigma$ clipping removed 3.9\% of the data points from {the} 2014 and 
1.4\% from the 2015 data sets. This new high-pass filtered time series
was then used to recalculate the variances,~$\sigma_i^2$, and weights were
assigned to each data point as:
\begin{equation}
  w_{i} = \frac{1}{\sigma_i^2}.
\end{equation}
The median rms noise is 1.47\,m\,s$^{-1}$ and the best observing periods have
noise levels below 1.3\,m\,s$^{-1}$ (16\% of the data points).  The noise levels
are above 2\,m\,s$^{-1}$ for only 6\% of the data points.

\begin{figure}[tbh]
\epsscale{1.00}
\plotone{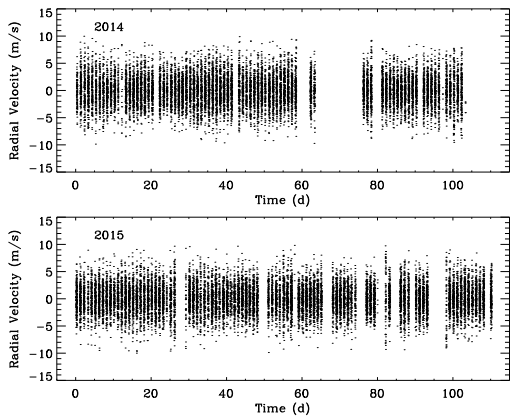}
\caption{Times series for $\mu\,$Her after 4-$\sigma$ clipping for the
  observations from 2014 (top) and 2015 (bottom panel). }
\label{fig:time-series-all}
\end{figure}

After removing the bad data and assigning statistical
weights to each data point, we calculated the power spectrum,  
as described in the next section.  The full time series is shown in 
Fig.~\ref{fig:time-series-all}. Note that the nightly average was 
subtracted from each night, which removes long--period variations but does 
not affect the oscillation signal.

\section{Data analysis} \label{sec:oscillations}

\subsection{Calculating the power spectrum} \label{subsec:powerspectrum}

\begin{figure}[tbp]
\epsscale{1.00}
\plotone{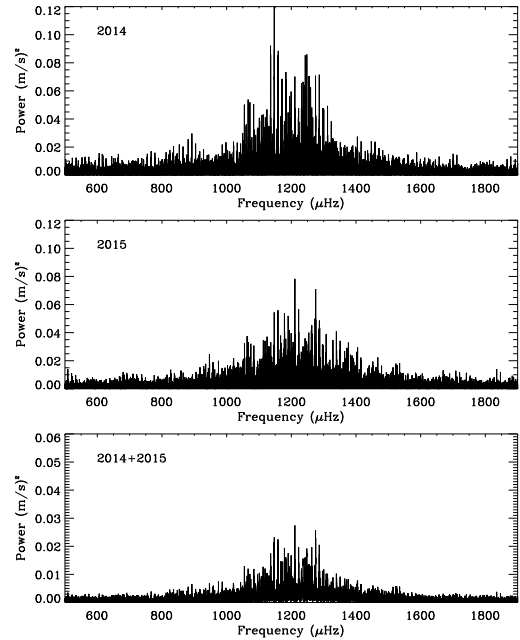}
\caption{The power spectra for the individual series from 2014  
         and 2015 data as well as the combined spectrum. For details see text.}
\label{fig:powerspectrum}
\end{figure}

The power spectrum of the $\mu\,$Her time series was calculated as a weighted 
fit of sinusoids, following the algorithms described by \cite{Frandsen1995} 
and \cite{Handberg2013}.  We calculated power spectra separately for the 
2014 and 2015 series, and then combined them into one power spectrum as a 
weighted average based on their mean noise levels.  The relative weights 
were 42\% and 58\% for the 2014 and 2015 data, respectively.  The individual 
and combined power spectra are shown in Fig.~\ref{fig:powerspectrum}.

The noise level in the combined power spectrum corresponds to 2\,cm\,s$^{-1}$ 
in amplitude at a frequency of 3000\,$\mu$Hz, which translates to 
19.4\,${\rm cm}^2{\rm s}^{-2}\mu$Hz$^{-1}$ in power density. 
This is similar to the noise 
levels in the $\alpha$~Cen A and B time series data 
\citep[see][]{Butler2004, Bedding2004, Kjeldsen2005}.  For example, the noise 
level in amplitude for $\alpha$~Cen B was 1.4 cm\,s$^{-1}$ at 7000\,$\mu$Hz, 
but close to 2 cm\,s$^{-1}$ at 3000\,$\mu$Hz.  Thus the 1-m SONG telescope and 
spectrograph has achieved a noise level in $\mu\,$Her over the 200 nights that 
is comparable to that achieved with the 8-m VLT and 4-m AAT over nine nights 
in a star that is 7 times brighter.

\begin{figure}[tbp]
\epsscale{1.00}
\plotone{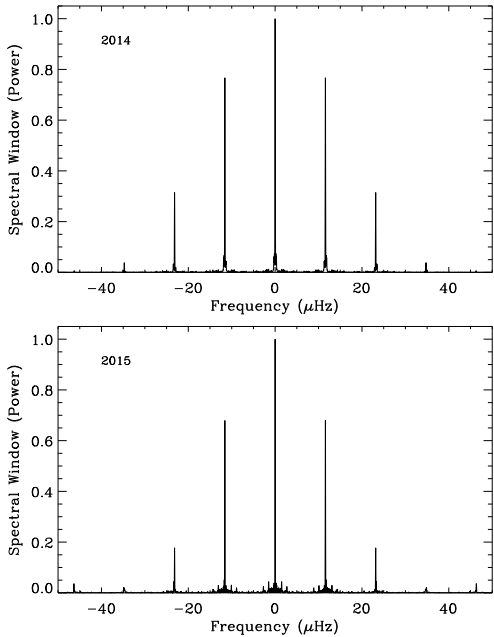} 
\caption{The spectral windows for the 2014 and 2015 data.}
\label{fig:window-function}
\end{figure}

\begin{figure}[tbp]
\epsscale{1.00}
\plotone{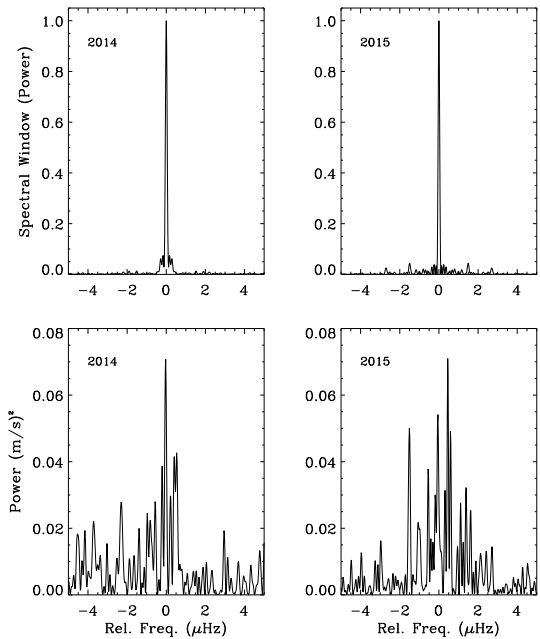}
\caption{ Upper panels: Close-ups of the spectral windows for the 2014 and 2015 data.
Lower panels: Close-up view of the frequency peak at 1274.93\,$\mu$Hz showing the spread of
oscillation power caused by the stocastic nature of the oscillations. }
\label{fig:window-function-zoom}
\end{figure}

Extraction of mode frequencies was done in the combined power spectrum. A
large number of p-modes are clearly present, especially near
the maximum power at 1200\,$\mu$Hz. However, the single-site data result in a
complicated spectral window with strong sidelobes.  The spectral windows
for the 2014 and 2015 data are shown in Figs.~\ref{fig:window-function}
and \ref{fig:window-function-zoom}.  


\subsection{Identifying the p-modes} \label{subsec:p-modes}

\begin{figure}[tbp]
\epsscale{1.0}
\plotone{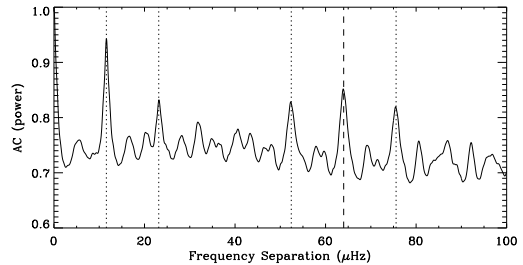}
\caption{The autocorrelation of the power spectrum smoothed using a
Gaussian with a FWHM of 0.5\,$\mu$Hz.  The  dashed
  line indicates the peak at 64\,$\mu$Hz that we identify as the large
  separation, together with extra peaks (dotted lines) corresponding to
  daily sidelobes in the spectral window.}
\label{fig:auto-correlation}
\end{figure}

As a next step we determined the large frequency separation.  
Figure~\ref{fig:auto-correlation} shows the
autocorrelation of the power spectrum after smoothing with a Gaussian
of FWHM 0.5{\AA}, for frequency shifts
between 0 and 100\,$\mu$Hz.  The peaks at 11.6 and 23.1\,$\mu$Hz correspond
to 1 and 2 cycles per day, respectively, arising from the daily gaps.  
We can identify the large frequency separation of $\mu\,$Her as 
$\Delta\nu = 64\,\mu$Hz. 
This value for $\Delta\nu$ agrees with the prediction by \cite{Bedding1996}, which 
was based on their estimates of the mass and radius of
the star.  It is also consistent with the observed value of $\nu_{\rm max} =
1200\,\mu$Hz. Our measurement disagrees
with the value of 56.5\,$\mu$Hz determined by \cite{Bonanno2008} based 
on 7 nights of radial-velocity measurements.  However, we note that in their 
Fig.\,3, which is a comb-response function of their power spectrum (analogous to 
an autocorrelation), there is a secondary peak close to 64\,$\mu$Hz. The incorrect 
determination of $\Delta\nu$ is most likely due to the short time span of the 
observations and the confusion with the daily sidebands. 

\begin{figure*}[tbh]
\epsscale{1.0}
\plottwo{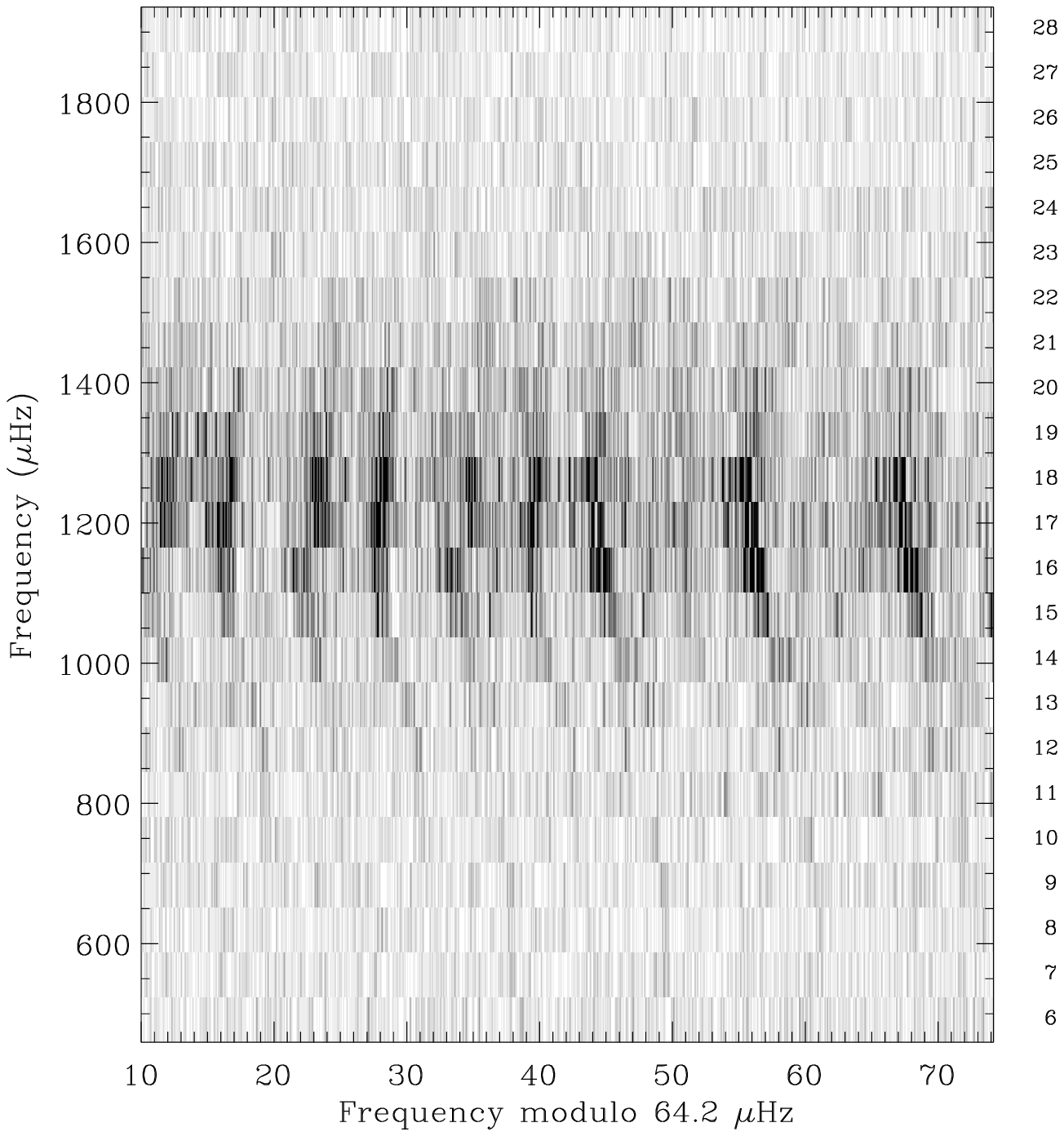}{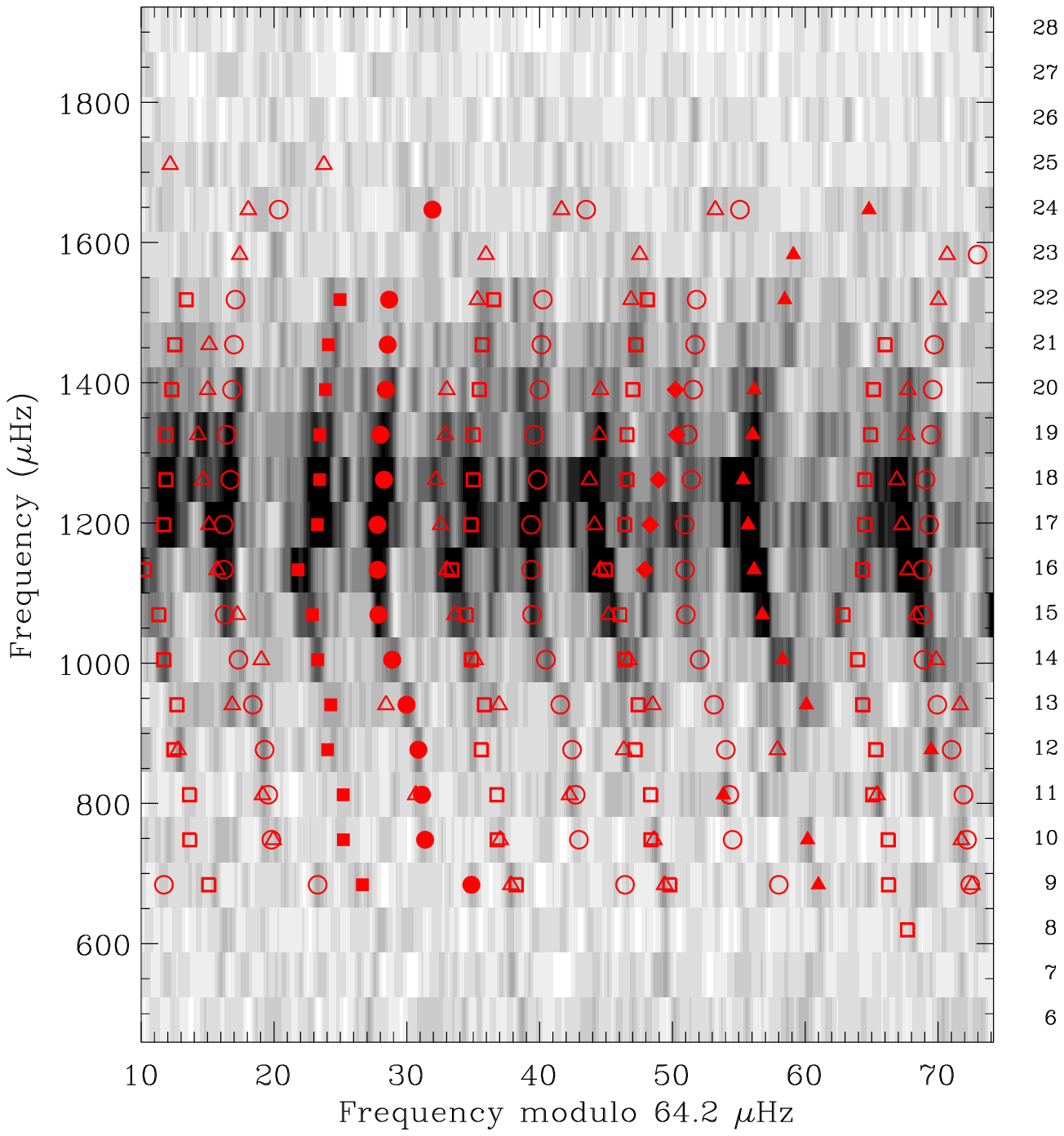}
\caption{The observed power spectrum of $\mu\,$Her in \'echelle 
 format as a grayscale image, with no smoothing (left) and smoothed to 
 a resolution of 0.5\,$\mu$Hz (right).  In the right plot, the filled 
 symbols show the 49 possible modes that we have identified (see text).  
 The open symbols show the first and second daily sidelobes on each 
 side of these modes.  Symbol type indicates the mode degree: $l=0$ (circles), 
 $l=1$ (triangles), $l=2$ (squares) and $l=3$ (diamonds). The numbers on 
 the right of each plot show the radial order, which corresponds to $n$ for the 
 $l=0$ modes.  }
\label{fig:greyscale-echelle}
\end{figure*}

\begin{figure}[tbh]
\epsscale{1.0}
\plotone{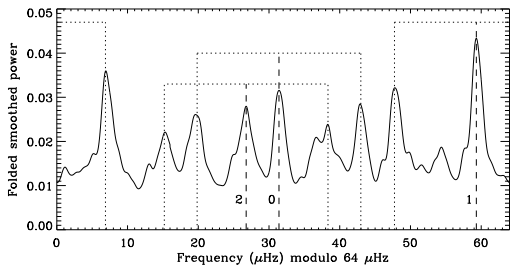}
\caption{Folded smoothed power spectrum for the frequency range 976-1424\,$\mu$Hz. 
The peaks corresponds to $l$ = 0, 1 and 2 shown by long dashed lines as well 
as the spectral window (1/d) shown by dotted lines.}
\label{fig:folded-power}
\end{figure}

Figure~\ref{fig:greyscale-echelle} illustrates the observed power 
spectrum in \'echelle format, where we see clear vertical ridges corresponding 
to modes with different degrees. 
We tested other values of $\Delta\nu$ and found that none gave the same
clear structure of vertical ridges.
Due to the single-site nature of the data, the first and second daily sidelobes are prominent.
To locate the individual oscillation modes we constructed a
folded power spectrum using the following procedure. We first smoothed the power spectrum by using
a Gaussian function with a FWHM of 1\,$\mu$Hz. This was then folded with a
spacing of 64\,$\mu$Hz between 976\,$\mu$Hz and 1424\,$\mu$Hz (7 radial orders,
centered at 1200\,$\mu$Hz). The resulting folded power spectrum is shown in
Fig.~\ref{fig:folded-power}, where the positions of
modes of different degrees ($l$ = 0, 1 and 2) can be seen. 
We used the peaks identified in Fig.~\ref{fig:folded-power} to estimate the
parameters in the asymptotic relation \citep{Tassoul1980, Scherrer1983, Christ1988}:
\begin{equation}
 \nu(n,l) \approx \Delta\nu(n +  \textstyle{\frac{1}{2}} l +\epsilon) - l (l + 1)  D_0. \label{eq:asymptotic}
\end{equation}
We found $\Delta\nu = 64.2\,\mu$Hz, $D_0$ = 0.80 $\mu$Hz and
$\epsilon = 1.44$.  Note that this value of $\epsilon$ is consistent with
expectations for a star with the effective temperature of $\mu\,$Her 
\citep{White2012}.

Using Eq.~\ref{eq:asymptotic} we estimated the expected frequencies of the
individual p-modes and identified them in the power spectrum. Thanks to the 
high data quality, we also detected five $l=3$ modes in the range 
1100--1400\,$\mu$Hz, where the S/N is highest. Additionally, one bumped 
$l=1$ mixed mode is apparently present at low frequencies.

In total, we identified 49 probable modes, shown as filled symbols
in the right panel of Fig.~\ref{fig:greyscale-echelle}, superimposed on a
smoothed version of the observed power spectrum.  The open symbols show the
first and second daily sidelobes on either side of each mode.  It is
remarkable that none of the sidelobes coincide with other p-modes or their
sidelobes.  It is extremely fortunate that the single-site spectral window
has little impact on our efforts to identify and measure the
oscillation modes.  
Indeed, it seems that its frequency spacings make $\mu\,$Her an ideal 
target for single-site observations \citep[see][for a discussion of SONG's spectral 
window and its influence on choice of targets]{Arentoft2014}.

\begin{figure*}[tbh]
\epsscale{1.00}
\plotone{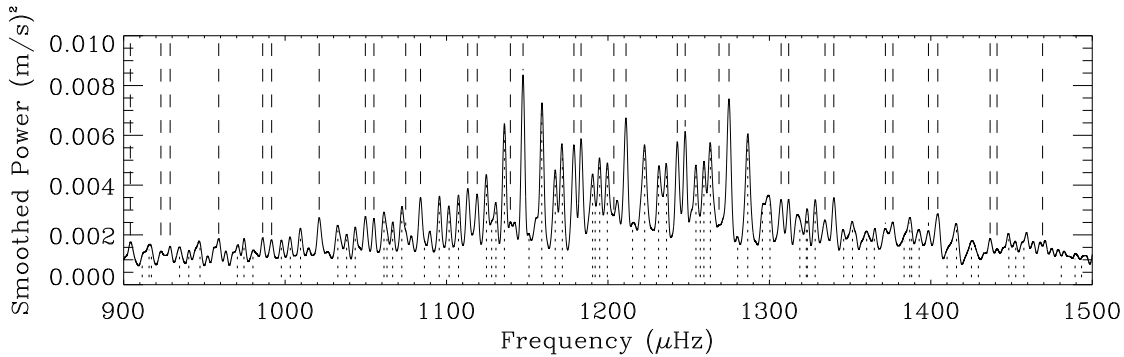} 
\caption{Central part of the combined power spectrum for $\mu\,$Her,
  smoothed with a Gaussian having FWHM of 3\,$\mu$Hz to enhance the
  visibility of the peaks. Dashed lines above the 
  smoothed power spectrum show the peaks identified to be oscillation modes; 
  the dotted lines below illustrate their daily sidelobes.}
\label{fig:smoothed-power-2}
\end{figure*}

We estimated uncertainties in the frequencies based on their S/N using
a procedure similar to \citet[][Section 4]{Kjeldsen2005}.  
These frequencies and their uncertainties were used as input for the 
Markov Chain Monte Carlo (MCMC)
frequency extraction described in Sec.~\ref{subsec:mcmc}.  The identified modes
in the central part of the spectrum, together with their daily sidelobes,
are shown in Fig.~\ref{fig:smoothed-power-2}.

\subsection{Amplitude and frequency of maximum power  } \label{subsec:amplitude}

To determine the frequency of maximum power ($\nu_{\rm max}$) and the peak
oscillation amplitude ($A_{\rm osc}$) for $\mu$\,Her, we followed the
procedure described in Section 3.2 of \citet{Kjeldsen2008}. 
This involves smoothing the power spectrum to estimate the total power 
in the oscillations in a manner that is insensitive to the spectral window. 

We found
the following values: $\nu_{\rm max} = 1216 \pm 11\,\mu$Hz and $A_{\rm osc}
= 38.9 \pm 1.2\,{\rm cm\,s}^{-1}$.  Note that this velocity amplitude
corresponds to radial modes and is 2.08 $\pm$ 0.10 times the mean solar
value (see \citealt{Kjeldsen2008} for details).

Interestingly, the oscillation amplitude for $\mu$\,Her decreased
significantly from 2014 to 2015.  This is clearly seen in
Fig.~\ref{fig:powerspectrum}.  Analyzing the two power spectra separately,
as described above, showed the peak amplitude to be $41.6 \pm 1.7\,{\rm
cm\,s}^{-1}$ in 2014 and $36.1 \pm 1.5\,{\rm cm\,s}^{-1}$ in 2015.


\subsection{Extraction of mode properties using MCMC analysis  } \label{subsec:mcmc}

The next step was to measure parameters for the 49 individual modes
using an MCMC analysis.
In order to use the full timespan of the measurements 
we constructed a full time series using all the available data from 
the two observing runs. 
The two time series were concatenated, but the gap between them was reduced
to 80 days.  This can be justified by the fact 
that we are searching for stochastic oscillations where the mode-lifetime is 
significantly shorter than 80 days. In this way we ensure that any 
oscillations from the 2014 data set will have disappeared and do not affect 
the 2015 data set. This concatenation creates a better 
window function.
The final power density spectrum and corresponding spectral window function were then
calculated from the time series specified above, following the prescriptions outlined in 
Section~\ref{sec:processing-time-series}.


\subsubsection{MCMC peakbagging}
The fit to the power spectrum was performed using the APT~MCMC 
algorithm \citep{Handberg2011} and the preliminary frequencies 
determined in Section~\ref{subsec:p-modes} were used as starting guesses.
We ran 4 million iterations, which were subsequently 
thinned to 2 million, using 10 parallel tempering levels to avoid local 
maxima solutions. The model limit spectrum that was fitted to the observed power 
spectrum was defined as:

\begin{equation}
   \mathscr{P}(\nu) = \eta(\nu) \sum_{n,l} \sum_{m=-l}^l \frac{H_{nl} \mathscr{E}_{l m}(i)}{1 + \frac{4}{\Gamma_{nl}^2} (\nu - \nu_{nl} - m\delta\nu_s)^2} + N(\nu) \,
\end{equation}
where $\nu_{nl}$ is the mode frequency, $H_{nl}$ is the mode 
height, $\Gamma_{nl}$ is the linewidth (which is inversely proportional to 
the mode lifetime) and $\delta\nu_s$ is the rotational splitting. 
The factor $\eta(\nu)\equiv\sinc^2(\Delta T_\mathrm{int} \nu)$ is the attenuation of signals 
arising from the non-zero integration time ($\Delta T_\mathrm{int}$). The noise 
model $N(\nu)$ was simply a white--noise profile across the region of interest.
In order to limit the number of free parameters, $H_{nl}$ 
and $\Gamma_{nl}$ were linearly interpolated in frequency 
between $H_{n0}$ and $\Gamma_{n0}$, respectively, and the height was scaled with 
the visibility of the mode \citep[see][]{Handberg2011}. The relative heights 
of rotationally split components within a multiplet were taken
as \citep{Gizon2003}:

\begin{equation}
  \mathscr{E}_{l m}(i) = \frac{(l - |m|)!}{(l + |m|)!} \lbrace P_{l}^{|m|}(\cos i) \rbrace^2 \,
\end{equation}
where $P_l^m(x)$ are the associated Legendre functions.

Instead of using the mode height, $H_{nl}$, directly as the free parameter 
in the fit, the mode amplitude was used. This is less correlated with 
the linewidth, $\Gamma_{nl}$, and therefore provides a more stable fit. 
The conversion from amplitude to height was done following \citet{Fletcher2006}, 
which allows for linewidths becoming comparable to the frequency resolution. 
Similarly, the projected rotational splitting, $\nu_s\sin i$, was used as the 
free parameter instead of the rotational splitting itself, to avoid known correlations.

Uniform priors were set for mode frequencies, the rotational splitting and 
the inclination angle,  
whereas modified Jeffreys priors were used for mode heights and linewidths.

In order to account for the single-site window function,
the model spectrum, ${P}(\nu)$, was convolved with the spectral 
window in each iteration of the MCMC. This has a very significant 
impact on the computing time, but is essential in order to describe the spread 
of power to sidelobes due to the non-continuous observations.

\textcolor{black}{From the resulting Markov chain, the final parameters and errors listed in 
Table~\ref{tab:2} were estimated from the full posterior probability 
distributions as the median values and 68.3\% confidence interval}. The final 
mode frequencies (with uncertainties) were corrected for the systemic radial-velocity Doppler 
shift ($v_\mathrm{rad} = -17.07{\pm}0.12$\,km\,s$^{-1}$) in order to list the frequencies 
in the rest frame of the star \citep{Davies2014}.

Finally, we calculated the frequency-separation ratio as defined by \citet{Roxburgh2003},
which are used in the following sections for modeling of the observations:

\begin{align}
    r_{01}(n) &= \frac{1}{8} \frac{\nu_{n-1,0} - 4\nu_{n-1,1} + 6\nu_{n,0} - 4\nu_{n,1} + \nu_{n+1,0}}{\nu_{n,1} - \nu_{n-1,1}} \\
    r_{02}(n) &= \frac{\nu_{n,0} - \nu_{n-1,2}}{\nu_{n,1} - \nu_{n-1,1}} \\
    r_{10}(n) &= \frac{-1}{8} \frac{\nu_{n-1,1} - 4\nu_{n,0} + 6\nu_{n,1} - 4\nu_{n+1,0} + \nu_{n+1,1}}{\nu_{n+1,0} - \nu_{n,0}} \, .
\end{align}

These were calculated using the full Markov chains for each frequency coming 
from the MCMC analysis, yielding the full correlation matrices between all ratios.

\begin{figure}
\epsscale{1.00}
\plotone{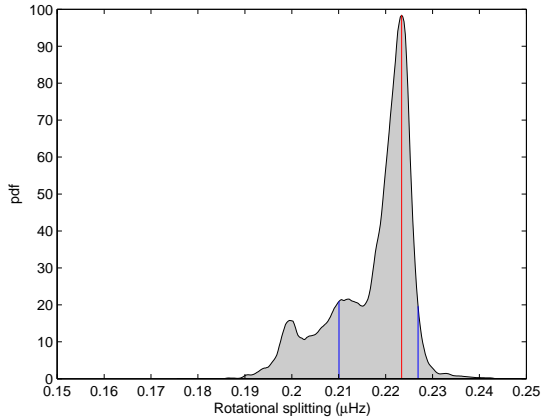}
\caption{Probability density function (pdf) for the rotational frequency splitting from 
MCMC analysis. The red vertical line indicates the mode of the posterior distribution, 
and the two blue vertical lines show the 68.3\% highest probability density region.}
\label{fig:mcmc_rotation}
\end{figure}

\begin{figure}
\epsscale{1.00}
\plotone{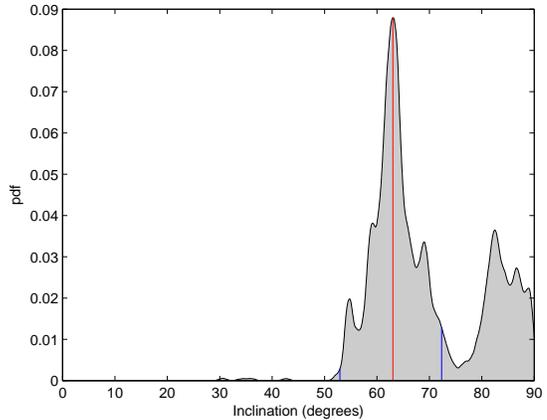}
\caption{Probability density function (pdf) for the inclination angle of  $\mu$\,Her
         from MCMC analysis. The red vertical line indicates the mode of the 
         posterior distribution, and the two blue vertical lines show the 
         68.3\% highest probability density region.}
\label{fig:mcmc_inclination} 
\end{figure}

From the MCMC analysis, we were also able to constrain the rotational splitting 
between the different $m$-components of the $l=1$ multiplets. We also measured the stellar inclination 
angle, based on the relative heights of these $m$-components
 \citep{Gizon2003} (see Figs.~\ref{fig:mcmc_rotation} and \ref{fig:mcmc_inclination}).
The resulting rotational period is $P_\mathrm{rot} = 52^{+3}_{-1}$ days and the 
stellar rotational inclination angle is $i = 63^{+9}_{-10}$ degrees. 
Both values and their errors were determined as the mode values and 68.3\% confidence
intervals in Fig.~\ref{fig:mcmc_rotation} and Fig.~\ref{fig:mcmc_inclination}. 

%
%

\begin{table*}[h]
\begin{center}
\caption{Frequencies ($\mu$Hz) for individual oscillation modes extracted from the
  MCMC analysis, listed in \'echelle format (see
  Fig.~\ref{fig:greyscale-echelle}).  Note that $n$ corresponds to the
  radial order for the $l=0$ modes.  \label{tab:2}}
\begin{tabular}{ccccc}
\tableline\tableline

$n$ & {$l=2$}                   & {$l=0$}                   & {$l=3$}                   & {$l=1$}                   \\
    &                                                                                         \\
\tableline
 24 & $1636.68^{+0.30}_{-0.45}$ &  ---                      &  ---                       & $1669.55^{+0.27}_{-0.37}$ \\
 23 &  ---                      &  ---                      &  ---                       & $1599.67^{+0.68}_{-0.22}$ \\
 22 & $1501.34^{+0.20}_{-0.42}$ & $1505.04^{+0.21}_{-0.72}$ &  ---                       & $1534.83^{+0.72}_{-0.30}$ \\
 21 & $1436.28^{+0.05}_{-0.10}$ & $1440.74^{+0.02}_{-0.05}$ &  ---                       &  ---                      \\
 20 & $1371.87^{+0.24}_{-0.33}$ & $1376.41^{+0.10}_{-0.14}$ &  $1398.21^{+0.53}_{-0.43}$ & $1404.15^{+0.12}_{-0.17}$ \\
 19 & $1307.24^{+0.13}_{-0.12}$ & $1311.81^{+0.08}_{-0.13}$ &  $1334.13^{+0.52}_{-0.25}$ & $1339.85^{+0.10}_{-0.10}$ \\
 18 & $1243.05^{+0.07}_{-0.06}$ & $1247.89^{+0.04}_{-0.04}$ &  $1268.56^{+0.11}_{-0.27}$ & $1274.93^{+0.05}_{-0.06}$ \\
 17 & $1178.69^{+0.11}_{-0.09}$ & $1183.20^{+0.05}_{-0.05}$ &  $1203.74^{+0.37}_{-0.26}$ & $1211.12^{+0.05}_{-0.06}$ \\
 16 & $1113.04^{+0.05}_{-0.07}$ & $1119.03^{+0.06}_{-0.07}$ &  $1139.15^{+0.57}_{-0.18}$ & $1147.38^{+0.04}_{-0.04}$ \\
 15 & $1049.94^{+0.21}_{-0.11}$ & $1054.90^{+0.06}_{-0.06}$ &  ---                       & $1083.81^{+0.06}_{-0.06}$ \\
 14 &  $986.14^{+0.12}_{-0.09}$ &  $991.75^{+0.11}_{-0.14}$ &  ---                       & $1021.14^{+0.13}_{-0.15}$ \\
 13 &  $922.93^{+0.08}_{-0.12}$ &  $928.64^{+0.06}_{-0.06}$ &  ---                       &  $958.75^{+0.17}_{-0.17}$ \\
 12 &  $858.51^{+0.13}_{-0.04}$ &  $865.34^{+0.10}_{-0.12}$ &  ---                       &  $903.95^{+0.08}_{-0.07}$ \\
 11 &  $795.49^{+0.02}_{-0.45}$ &  $801.42^{+0.06}_{-0.49}$ &  ---                       &  $824.11^{+0.26}_{-0.03}$ \\
 10 &  $731.32^{+0.15}_{-0.15}$ &  $737.47^{+0.27}_{-0.21}$ &  ---                       &  $766.27^{+0.03}_{-0.02}$ \\
  9 &  $668.56^{+0.19}_{-0.05}$ &  $676.76^{+0.02}_{-0.03}$ &  ---                       &  $702.89^{+0.08}_{-0.08}$ \\
\tableline
\end{tabular}
\end{center}
\end{table*}

\section{Modeling the oscillations} \label{sec:modeling}

Once the oscillation frequencies were extracted we used different codes and 
procedures to model the observations of $\mu$\,Her.
In this section we provide detailed descriptions of this endeavor.
 
\subsection{Fits to individual frequencies} \label{subsec:individual}

\def\note #1]{{\bf #1]}}
\def\Msun{\,{\rm M}_\odot}
\def\Rsun{\,{\rm R}_\odot}
\def\Lsun{\,{\rm L}_\odot}

We fitted the 49 frequencies in Table~\ref{tab:2} and their ratios following 
procedures described by \citet{Silva2015}, in this case taking into account 
the presence of the mixed modes.
One fit was applied to the individual frequencies, which were assumed to 
be statistically independent.
A grid of models and oscillation frequencies was calculated using 
the ASTEC stellar evolution code \citep{Christ2008a} and the ADIPLS
adiabatic pulsation code \citep{Christ2008b}.
The evolution modeling used the OPAL equation of state
\citep{Rogers2002} and OPAL opacities \citep{Iglesi1996}, 
supplemented by the \citet{Fergus2005} low-temperature opacities.
The nuclear reaction rates were obtained from the NACRE compilation
\citep{Angulo1999}.
Diffusion and settling of helium and heavy elements were not included.
Convection was described using the mixing-length formalism
\citep{Bohm1958}, and convective overshoot was not included.
The grid spanned a large range in mass and composition, although constrained
by an assumed Galactic chemical evolution model with
$\Delta Y/\Delta Z = 1.4$, where $Y$ and $Z$ are the abundances of
helium and heavy elements, respectively.
Models with three values of the mixing-length parameter $\alpha_{\rm ML}$,
1.5, 1.8 and 2.1, were included, where $\alpha_{\rm ML} = 1.8$ roughly
corresponds to the solar calibration.

To match the observed frequencies, the computed frequencies were corrected
for the errors introduced by the treatment of the near-surface layers
by applying a fitted scaled solar surface correction \citep{Christ2012},
described in more detail, together with other aspects of this so-called ASTFIT
fitting technique, by \citet{Silva2015}. 
Briefly, the fit is carried out by minimizing, along each evolution sequence,

\begin{equation}
\chi^2 = \chi^2_{\rm spec} + \chi_\nu^2 \;. 
\label{eq:minim}
\end{equation}
Here, $\chi_{\rm spec}^2$ is based on observed values of $T_{\rm eff}$ 
and [Fe/H] (cf. Table~\ref{tab:muher-parameters}), and

\begin{equation}
\chi_\nu^2 = {1 \over N - 1} \sum_{i=1}^N 
\left( \nu_i^{\rm (obs)} - \nu_i^{{\rm(mod)}} \over \sigma_i \right)^2
\label{eq:chisqnu}
\end{equation}
is based on the observed frequencies $\nu_i^{\rm (obs)}$ and standard
deviations $\sigma_i$ listed in Table~\ref{tab:2}.
The remaining observed properties were not included in the fit but were 
used to check the results.
The model frequencies, $\nu_i^{\rm (mod)}$, included the surface correction (see above).
An initial minimization was carried out between timesteps 
in the evolution sequence by assuming that the frequencies scale 
as $R^{-3/2}$.
This defined a minimum $\chi_{\rm min}^2$ for each evolution track in the grid.
The best-fitting models were found by locating the smallest resulting values
of $\chi_{\rm min}^2$.

Owing to the presence of mixed modes, the $R^{-3/2}$ scaling of the frequencies
is not universally valid, leading to potential systematic errors in the fits.
To correct for this, the fit was refined by computing, in the vicinity of
the minima determined by the above scaling procedure, frequencies for a small
set of models suitably interpolated between timesteps in the evolution
sequence.  As shown by \citet{Christ2010}, this allows to fully resolve 
the behavior of the frequencies in the vicinity of an avoided crossing 
involving mixed modes.  In practice, this was applied only to evolution tracks 
where the $\chi_{\rm min}^2$ as determined by the simple procedure was less
than twice the minimum amongst the values of $\chi_{\rm min}^2$ so 
determined.

\begin{figure}
\epsscale{1.00}
\plotone{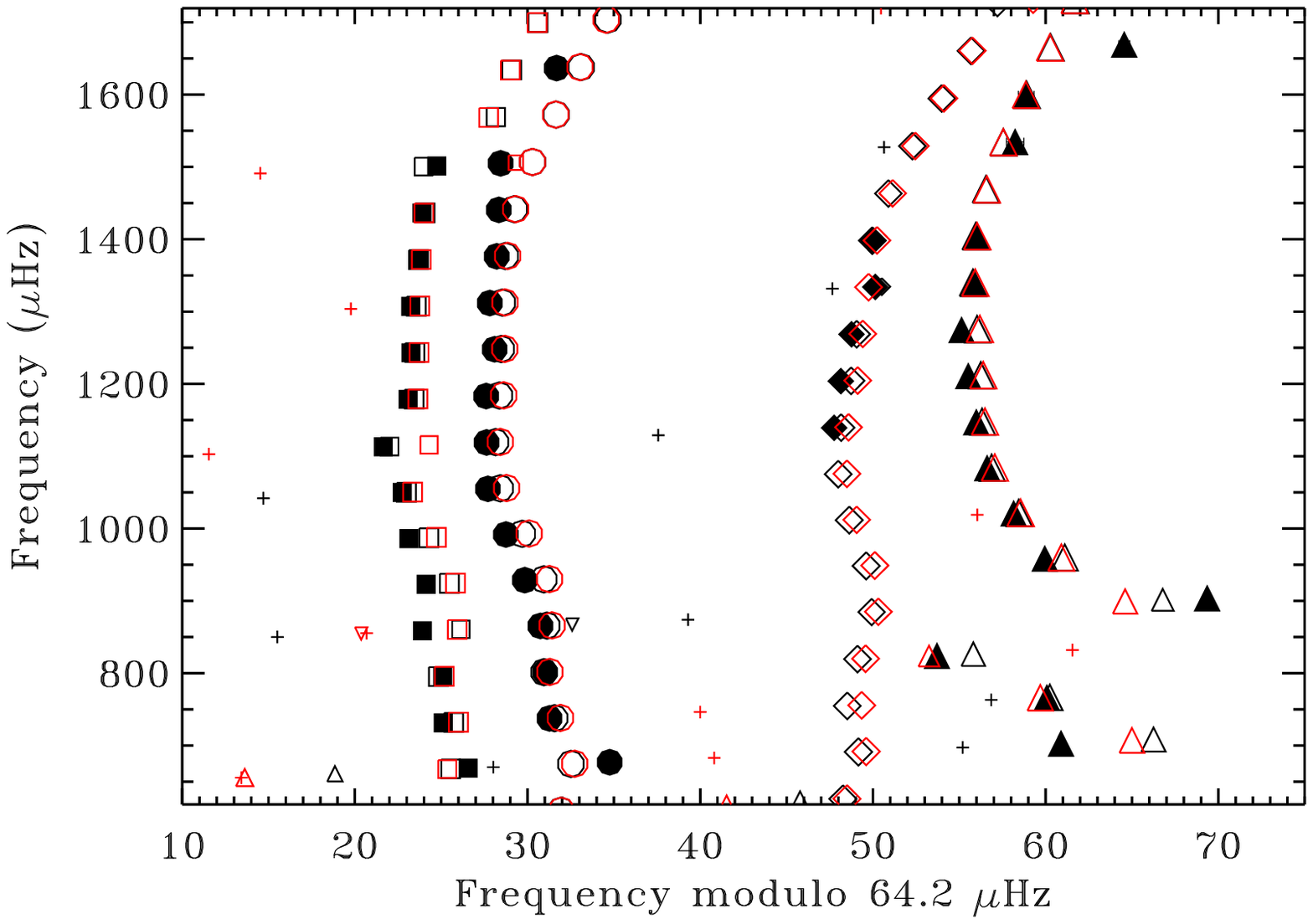}
\caption{ \'Echelle diagram of observed and fitted frequencies.
The filled black symbols show the frequencies provided in 
Table~\ref{tab:2}, while the black and red open symbols are for 
best-fitting models, after solar-scaled surface correction,
with masses of respectively $1.12$ and 1.15\,M$_\odot$ 
(model ASTFIT1 and ASTFIT2 in Table~\ref{tab:modres}).
Circles, triangles, squares and diamonds show results for
$l = 0, 1, 2$ and $3$, respectively. Symbol sizes are based 
on a rough estimate of mode amplitudes, relative to the neighboring 
radial mode \citep[see][]{Christ1995} For $l = 2$ and $3$ small symbols, 
corresponding to strongly mixed modes, have been replaced by plusses.  
Inverted triangles at a frequency near $860 \,\mu {\rm Hz}$ mark 
strongly mixed dipolar modes.
\label{fig:modechl}}
\end{figure}

The stars analyzed by \citet{Silva2015} were all on the main sequence
and the observed modes were purely acoustic.  In contrast, $\mu\,$Her 
is a subgiant with clearly identified mixed modes (Fig.~\ref{fig:modechl}),
and the relevant models also have several mixed modes.
This complicates the identification of the observed modes with
those of the models in the grid.
We have applied a relatively simple technique to identify the relevant
model modes in cases with mixed modes,
taking into account that the present observations show only one nonradial
mode of each degree in each interval between two adjacent radial modes.
Thus in each radial-mode interval we chose (with an exception noted below)
that frequency of a given degree which minimized the normalized inertia
\begin{equation}
Q_{nl} = {E_{nl} \over \bar E_0(\nu_{nl})} \; ,
\end{equation}
where $E_{nl}$ is the inertia of the mode and $\bar E_0(\nu_{nl})$
is the radial-mode inertia, interpolated logarithmically to the 
frequency $\nu_{nl}$ of the given mode.
The underlying assumption is that this is the mode most likely to be
observed.

For $l = 2$ and $3$ there was typically a clear minimum of $Q_{nl}$ 
amongst the relevant modes, and the above procedure produced a reasonable fit.
For $l = 1$, however, there may be two modes in a given radial-mode interval
with comparable values of $Q_{nl}$, and there is a risk that the selected
mode does not provide the optimal fit to the observations.
To circumvent this problem, the procedure was modified by including in
the minimization a suitably weighted measure of the distance to the 
nearest observed dipolar mode. Although fairly crude, this method
yielded a reasonable behavior of the fit along the evolution tracks.

\begin{figure}
\epsscale{1.00}
\plotone{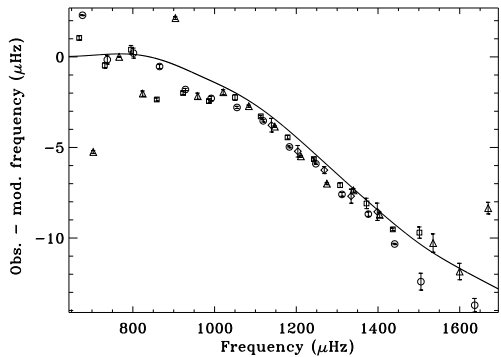}
\caption{Differences between observed frequencies (Table~\ref{tab:2})
and model frequencies, without surface correction, for the
$1.12 \Msun$ model ASTFIT1 in Table~\ref{tab:modres}.
The same symbols are used as in Fig.\,\ref{fig:modechl}.
The curve shows the scaled solar surface correction used in the fit.
\label{fig:freqdif}}
\end{figure}

As applied by \citet{Silva2015}, ASTFIT determined likelihood-weighted 
averages of the various stellar parameters.
In the present case, we have found that $\chi_\nu^2$
(cf. Eq. \ref{eq:chisqnu})
is dominated by a few modes, particularly the dipolar mode undergoing
avoided crossing, and hence this statistical procedure has little meaning
(see also Fig.~\ref{fig:freqdif}).
For this preliminary analysis we therefore simply considered a few 
examples of optimized fits for representative selected evolution tracks,
chosen to yield values of $T_{\rm eff}$ and [Fe/H] within $2 \sigma$
of the observed values and $\chi_\nu^2$ near its minimum value.
These are listed in Table~\ref{tab:modres}.
Two examples with masses of $1.12 \Msun$ and $1.15 \Msun$ 
are shown in the \'echelle diagram in Fig.\,\ref{fig:modechl}. 
Figure\,\ref{fig:freqdif} shows the resulting frequency differences 
for the $1.12 \Msun$ model, compared with the fitted surface function.

To determine the uncertainties in the stellar properties, we also 
fitted combinations of p-mode 
dominated frequencies using the BAyesian STellar Algorithm \citep[BASTA, see][]{Silva2015}. 
Briefly, this Bayesian approach relies on a large grid of stellar models to determine 
the probability density function of a given stellar property based on the 
fit to a set of  observational quantities. In this case, we considered the 
spectroscopic constraints $T_{\rm eff}$ and [Fe/H] and the 
frequency-separation ratios $r_{01}$ and $r_{10}$ above 1000~$\mu$Hz (to 
avoid the impact of the mixed modes in the fit) as the input parameters to 
be reproduced. We report in Table~\ref{tab:modres}, the median and the 16 and 84 
percentiles of the posterior probability density function. The results are 
in excellent agreement with those obtained with ASTFIT, as well as with the 
independent radius determination from interferometry.

Using the effective temperature, large frequency separation and [Fe/H] (see 
Table~\ref{tab:results_summary}) as inputs, we also calculated the stellar 
parameters using the Asteroseismology Made 
Easy \citep[AME][]{Lundkvist2014} grid-based method and found the values to be 
in agreement with those listed in Table~\ref{tab:modres}.

\begin{table*}[h]
\begin{center}
\caption{Results of model fits.
\smallskip\label{tab:modres}}
\begin{tabular}{ccccccc}
\tableline\tableline
Model & $M$         & $R$       & $L$       & $T_{\rm eff}$ & [Fe/H] & Age \\
      & [$\Msun$]   & [$\Rsun$] & [$\Lsun$] &    [K]        & [dex]  & [Gyr] \\
\tableline

ASTFIT1 & 1.12      &  1.71     &  2.7      &   5650        &  0.26  &  7.6  \\
ASTFIT2 & 1.15      &  1.73     &  2.6      &   5600        &  0.30  &  7.9  \\
BASTA   & $1.11^{+0.01}_{-0.01}$ &  $1.71^{+0.01}_{-0.02}$ &  $2.6^{+0.1}_{-0.1}$ & $5600^{+50}_{-50}$ & $0.21^{+0.06}_{-0.06}$ & $7.8^{+0.3}_{-0.4}$ \\

\tableline
From Table~\ref{tab:muher-parameters} &           & $1.73 \pm 0.02$ & $2.54 \pm 0.08$ & $5560 \pm 80$ & $0.28 \pm 0.07$ & \\ 
\tableline
\end{tabular}
\end{center}
\end{table*}


\subsection{Modeling amplitudes and mode life times} \label{subsec:amp_models}

\begin{figure}[h!]
\epsscale{1.00}
\plotone{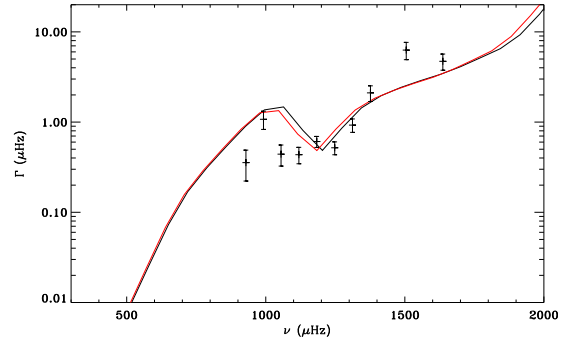}
\caption{Measured linewidths for radial modes from the MCMC analysis in Sec. \ref{subsec:mcmc}  
(full width at half maximum, symbols) are 
compared to theoretical estimates of twice the linear damping rates.
The black and red curves are the theoretical estimates 
from nonadiabatic stability analysis adopted for the global parameters 
of the models ASTFIT1 and ASTFIT2 respectively (see Table~\ref{tab:modres}). 
Only the mixing-length parameter was modified between the two stability 
analyses to reproduce the same surface-convection-zone depths as in
the two stellar models. }
\label{fig:damping-rates}
\end{figure}

Mode linewidths and amplitudes can be used to test models of stellar 
structure and stability.  In particular, comparison between observations 
and models can be used to calibrate the parameters in the convection model used 
in the numerical stability analysis.
In Fig.~\ref{fig:damping-rates} theoretical estimates of linear 
damping rates\footnote{The mode lifetime $\tau$ and the linewidths $\Gamma$ are 
related through $\tau=1/\pi \Gamma$.} of radial modes are compared to the SONG 
observations. 
These computations were performed for global model parameters of the models 
ASTFIT1 and ASTFIT2 (see Table~\ref{tab:modres}).

The depth of the (surface) convection zone was calibrated  
to the values obtained from the seismic models ASTFIT1 and ASTFIT2 described in 
Section~\ref{subsec:individual}. 
The basic stability computations were as in \cite{Houdek2006} using 
Gough's \citep{Gough77a, Gough77b} nonlocal, time-dependent convection model, but 
adopted for the stellar atmosphere a temperature -- optical depth ($T-\tau$) 
relation from \cite{Trampedach2014} 3D hydrodynamical simulations.  
The agreement with the observations is reasonably good. 

We also estimated the maximum value of the velocity amplitudes of 
the  acoustic oscillations.  Various excitation models have been 
used in the past to estimate amplitudes of stochastically excited 
oscillations \citep{Goldreich1977, Balmforth1992, Samadi2001, 
Houdek2006, Chaplin2005}. Here we adapt the scaling relation 
by \cite{Chaplin2011} for estimating the maximum velocity amplitude. 
Adopting the global parameters listed in 
Table~\ref{tab:muher-parameters}, we estimate for $\mu\,$Her a relative 
maximum velocity amplitude $V/V_\odot\simeq 1.83$ ($V_\odot$ being 
the maximum solar velocity amplitude), which is in reasonable 
agreement with the observed value of 2.08$\pm$0.10 discussed in 
Section~\ref{subsec:amplitude}.  We should, however, note that 
the $\beta$ function in \cite{Chaplin2011} equation (7) is rather 
uncertain and will add to the uncertainty from the adopted 
effective temperature for $\mu\,$Her.  The predicted 
value of the velocity amplitude will therefore capture the uncertainties 
in both the observations and the scaling relation.


\subsection{Using acoustic glitches}\label{subsec:glitches}

\begin{figure}[h!]
\epsscale{1.00}
\plotone{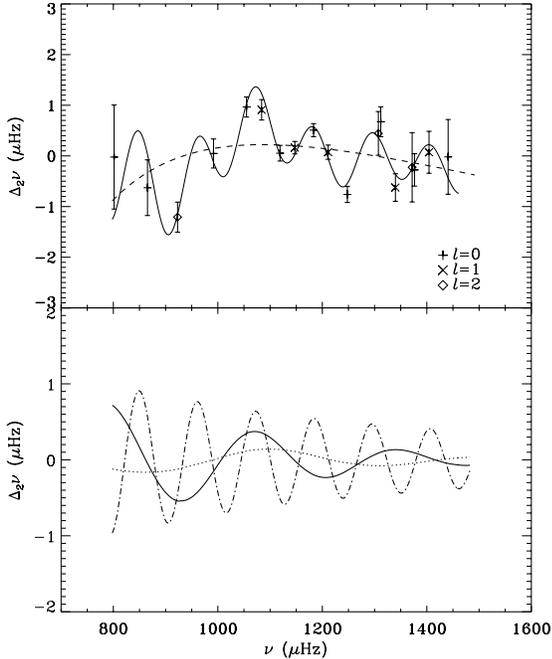}
\caption{Top: The symbols are second differences 
        $\Delta_2\nu_{n,l}:=\nu_{n-1,l}-2\nu_{n,l}+\nu_{n+1,l}$
        of low-degree ($l=0,1,2$) frequencies obtained from SONG. The 
        vertical bars represent standard errors, evaluated under the 
        assumption that the errors in the raw frequencies are independent. 
        The solid curve is the seismic diagnostic $D_2$ from \cite{Houdek2007}, 
        determined from fitting by least-squares the seismic diagnostic 
        to $\Delta_2\nu_{n,l}$. The dashed curve represents the smooth 
        contributions from the hydrogen ionization zones and super-adiabatic 
        layer.  Bottom: Individual contributions of the seismic diagnostic. 
        The solid curve is the contribution of the second stage of helium 
        ionization, the dotted curve displays the first helium ionization 
        contribution and the dot-dashed curve is the contribution from 
        the base of the convection zone.}
\label{fig:second-differences}
\end{figure}

Abrupt variations in the 
sound speed, which are called acoustic glitches, produce seismic 
signatures in the spacing of the observed frequencies. 
From these seismic signatures, the locations of the abrupt variation 
(in terms of acoustic depth $\tau$) can be estimated.
Figure~\ref{fig:second-differences} displays observed second differences 
$\Delta_2\nu_{n,l}:=\nu_{n-1,l}-2\nu_{n,l}+\nu_{n+1,l}$ 
of low-degree ($l=0,1,2$, symbols), together with results of 
the seismic diagnostic $D_2$ by \cite{Houdek2007}. 
This analysis adopts Airy functions for the pulsation eigenfunctions and 
glitches of both stages of helium ionization. We estimated the 
acoustic depths of the glitches brought about by the second stage 
of helium ionization, $\tau_{\rm II}$, and by the abrupt variation 
of the sound speed at the base of convection zone, $\tau_{\rm c}$. 
We found $\tau_{\rm II}\simeq1938\,$s and $\tau_{\rm c}\simeq4488\,$s. 
The depths of the acoustic glitches obtained directly from the equilibrium 
structures of the models listed in Table~\ref{tab:modres}, 
agree with $\tau_{\rm II}$ to within 3\% and for $\tau_{\rm c}$ to within 15\%. 
For the present work, we did not perform an error analysis for the 
acoustic-glitch depths, but plan to conduct a Monte-Carlo error analysis 
in an upcoming paper.

\section{Discussion and outlook} \label{sec:discussion}

Our 215 nights of observations of $\mu\,$Her represent the longest 
ground-based asteroseismology campaign of a solar-like star. For this
first SONG long-term target, we have determined
all the classical seismic observables and identified 49 oscillation
modes (see Table~\ref{tab:2}). 
\textcolor{black}{Using MCMC modeling we measured frequencies and linewidths 
of radial modes and rotational splitting of $l=1$ modes.} 
From this, the rotation period and inclination of the 
rotation axis were determined to be  $P_\mathrm{rot} = 52^{+3}_{-1}$ days 
and  $i = 63^{+9}_{-10}$ degrees (68.3\% confidence intervals), respectively.   
The observed oscillation frequencies were used as input for 
modeling and, taking into account 
the detected mixed mode, resulted in accurate 
determinations of the age as well as radius and luminosity in agreement 
with the observations (Table~\ref{tab:modres}).

\begin{table}[h]
\begin{center}
\caption{Summary of results  \smallskip\label{tab:results_summary}}
\begin{tabular}{lcl}
\tableline

Parameter                       &  Value               & Comment \\
\tableline
T$_{\rm eff}$[K]                &  5560$\pm$80         & J15    \\
$\log g$ [cgs]                  &  3.98$\pm$10         & J15    \\
${\rm [Fe/H]}$                  &  0.28$\pm$07         & J15    \\
$v\,\sin i$\,[km\,s$^{-1}$]     &  1.7$\pm$0.4         & J15    \\
\tableline
$\theta_{\rm LD}$[mas]          &  1.93$\pm$0.03       & measured \\
$R$ [$\Rsun$]                   &  1.73$\pm$0.02       & angular diameter + parallax \\
$L$ [$\Lsun$]                   &  2.54$\pm$0.08       & assuming $A_V\,=\,0$ \\
\tableline
$\nu_{\rm max}$ [$\mu{\rm Hz}$] & 1216$\pm$11          &  measured \\
$\Delta\nu$ [$\mu{\rm Hz}$]     & 64.2$\pm$0.2         &  measured \\
$\epsilon$                      & 1.44                 &  measured \\
$i$ [$^\circ$]                  & $63^{+9}_{-10}$      &  measured \\
P$_{\rm rot}$ [d]               & $52^{+3}_{-1} $      &  measured \\
\tableline
age [Gyr]                       & $7.8^{+0.3}_{-0.4}$  & from model \\
$M$ [$\Msun$]                   & 1.11$\pm$0.01        & from model \\
$R$ [$\Rsun$]                   & 1.72$\pm$0.02        & from model \\
$L$ [$\Lsun$]                   & 2.6 $\pm$0.1         & from model \\
$\log g$ [cgs]                  & 4.01$\pm$0.01        & from model \\
$\tau_{II}$ [s]                 & 1938                 & from model \\
$\tau_{c}$ [s]                  & 4488                 & from model \\
\tableline

\end{tabular}
\end{center}
\end{table}

We also compared the observed linewidths with theoretical values and found 
good agreement at frequencies around and above $\nu_{\rm max}$. For lower 
frequencies the mode lifetimes are significantly longer and our data were 
insufficient to resolve them. With the large number
of identified modes and the very good frequency precision, we determined 
the second frequency differences to measure the acoustic glitches 
associated with the \ion{He}{2} ionization layer and the base of the 
convection zone. Table~\ref{tab:results_summary} provides a full summary of our 
results. 

$\mu$\,Her is a very interesting seismic target, not only because it is 
ideal for single-site SONG observations but for several other reasons.  
For example, $\mu\,$Her and $\alpha\,$Cen\,A, the best asteroseismically 
studied bright solar-type stars, have the same mass to within the 
measurement uncertainties.  Very recently \cite{Pourbaix2016} redetermined 
the mass of $\alpha\,$Cen\,A to be $1.133\pm0.005$\,M$_\odot$ as compared 
to the 1.11-1.15\,M$_\odot$ reported here for $\mu$\,Her.  Note that this 
mass range is where the transition between convective and non-convective 
core on the main sequence occurs \citep[see][for an in-depth discussion of
$\alpha\,$Cen\,A]{Bazot2016}. 
Within 0.1\,dex, their reported metallicities are also identical. 
Thus, both stars should be on almost the same evolutionary track, allowing 
us to undertake comparative studies. 

We also note that $\mu$\,Her (given the metallicity, mass and age 
reported here) closely resembles stars in the old open cluster NGC\,6791.  
Thus, differential studies can improve constraints on the helium mass 
fraction, $Y$, of $\mu$\,Her and NGC\,6791.  
If shifted to the reddening and distance of NGC\,6791 using the $E(B-V)$ and 
apparent distance modulus derived by \cite{Brogaard2012}, $\mu$\,Her sits right 
on the cluster subgiant branch (SGB) of the color-magnitude diagram, confirming 
the near-identical properties of $\mu$\,Her and the stars in NGC\,6791.
The relative spectroscopic  $T_{\rm eff}$ of $\mu\,$Her and SGB twins in the cluster 
can then be used to tightly constrain the cluster reddening.  This can 
lead to an improved estimate of $Y$ for NGC\,6791 through reanalysis of the 
cluster \citep{Brogaard2012}, which also allows a precise estimate of $Y$ 
for $\mu$\,Her under the assumption of a common helium-to-metal enrichment for stars.

With the results presented here we are now in the position to 
learn more about the stellar obliquity\footnote{the angle between 
the orbital angular momentum and the stellar spin} of the $\mu\,$Her system, 
which is a quadruple system as specified in Section\,\ref{subsec:binary}.
From our seismic analysis we determined the inclination of the stellar
rotation axis $i$ of $\mu\,$Her to be $63^{+9}_{-10}$ degrees. This 
is very close to the inclination angle of the orbital
plane of $\mu$\,Her and its closest orbiting component, which is determined to be 
$63\pm5^\circ$ \citep{Roberts2016}. 

Combining our measurement of $i$ with the orientation of the
orbit does not give us the complete information on the obliquity because
we do not know the projection of the stellar spin axis on the plane
of the sky. Nevertheless, given the good agreement between the inclination 
of the stellar rotation axis and the orbital plane, we assume in the following that the
rotation axis of $\mu$\,Her is indeed aligned with the angular momentum of its orbit
and briefly discuss this finding.  There are only a handful of obliquity 
measurements in double stars \citep[see][for a list]{Albrecht2011}
that have a  short period (less then one month).
Among those, misaligned as well as
aligned systems were reported \citep[e.g.][]{ Albrecht2009, Albrecht2014, 
Albrecht2007, Triaud2013}.  For systems with larger semi-major axes, 
\citet[][and references therein]{Hale1994} estimated the stellar 
inclinations using the projected stellar rotation velocities ($v \sin i$) and 
found low obliquities in double-star systems with semi-major 
axes up to $\approx40$~AU. However, for systems with more than
two stellar components no indication of coplanarity was found. 
This was interpreted as a sign of long-term secular interactions 
(Kozai-cycles) between the different 
components \citep[e.g.][]{Fabrycky2007, Naoz2016, Anderson2016}. 
With an apparently low obliquity for the primary  component and a semi-major 
axis of  $2.9\pm0.3$~AU, the $\mu$\,Her system provides an interesting data 
point that does not seem to follow the trend observed by \cite{Hale1994}.

Based on 215 nights of observations we have presented the most 
detailed study of $\mu\,$Her to date but there is still much we can 
learn about this star. We will continue observing $\mu$\,Her during 
the coming years to improve the S/N in the power spectrum and the 
frequency resolution, and to check for oscillation amplitude and 
frequency variations.  We expect additional detections of low-frequency 
modes as well as more $l=3$ modes, which provide important constraints 
on the acoustic depth of the \ion{He}{2} ionization layer 
and the convection zone. Longer data sets, and therefore a higher S/N, 
will allow us to detect further mixed modes providing more insights 
about the deeper  regions of the star.   

With the upcoming NASA TESS (Transiting Exoplanet Survey Satellite) mission 
\citep{Ricker2015} we will be in the position to simultaneously observe 
$\mu$\,Her photometrically with TESS and spectroscopically with 
SONG\footnote{$\mu$\,Her has $R\,=\,2.9$ and $I\,=\,2.5$ which may be
too bright for TESS. We note that photometry on strongly saturated stars has
been done with success by White et al. (in prep.) for the Kepler mission data; 
hopefully this will be possible for TESS data as well.}. 
This will allow us to measure the oscillation amplitude ratios, providing detailed 
input on the convective properties of  $\mu$\,Her. The only other solar-type 
stars where similar observations were performed are the 
Sun \citep{Houdek2006, Jimenez2002} and Procyon \citep{Huber2011}. 

Based on its properties and the unprecedented data set, we expect $\mu$\,Her to become 
a benchmark star during the next years. With the addition of more SONG nodes, 
many of the brightest stars in the sky can be subject to similar 
comprehensive studies. Such work will provide a reference base of highly 
accurate parameters for the nearest stars, where the availability
of parallaxes and interferometric radii would provide 
strong model constraints. This will complement future space-based  
observations from TESS and PLATO.


\acknowledgments
\section*{Acknowledgements} \label{sec:acknowledgements}
We would like to acknowledge the Villum Foundation, The Danish 
Council for Independent Research $\mid$ Natural Science and the Carlsberg 
Foundation for the support on building the SONG prototype on Tenerife. 
The Stellar Astrophysics Centre is funded by The Danish National Research 
Foundation (Grant DNRF106) and research was supported by the ASTERISK project 
(ASTERoseismic Investigations with SONG and Kepler) funded by the European 
Research Council (Grant agreement n. 267864). We also gratefully acknowledge 
the support by the Spanish Ministry of Economy Competitiveness (MINECO) 
grant AYA2016-76378-P and the Australian Research Council. 
W.J.C. acknowledges support from the UK Science and Technology Facilities 
Council (STFC). M.S.L. is supported by The Danish Council for Independent 
Research's Sapere Aude program (Grant agreement no.:  DFF--5051-00130).
M.N.L. acknowledges the support of The Danish Council for Independent 
Research $\mid$ Natural Science (Grant DFF-4181-00415).
We thank the referee for a clear report resulting in improvements to
the material presented in this paper. 
A special thank you also goes to the staff at the Observatorio del Teide 
for their expert help during construction and operation of the telescope. 


\end{document}